%% file: iclr2026_conference.tex
\documentclass{article} %
\usepackage{iclr2026_conference,times}

\input{math_commands.tex}

\usepackage{hyperref}
\usepackage{url}
\usepackage{amsmath}
\usepackage{tikz}
\usetikzlibrary{arrows,positioning}

\title{Situation Perception: A Necessary Primitive to Artificial Superintelligence}

\author{Ziqin Yuan \& Jaymari Chua \\
School of Computer Science and Engineering \\
The University of New South Wales \\
Sydney, NSW 2052, Australia \\
}

\iclrfinalcopy %
\begin{document}

\maketitle

\begin{abstract}
Current large language models are extraordinary statistical engines. They compress vast amounts of text into useful patterns and can explain science, write code, imitate reasoning, and participate in philosophical conversation. Yet pattern mastery is not the same as general intelligence. A human infant begins with little explicit knowledge, but gradually discovers object permanence, cause and effect, other minds, bodily agency, and the persistence of the physical world. We make an argument that the path to artificial superintelligence (ASI) depends on a missing capacity we call \emph{situation perception}: the ability to construct, revise, and act within internal simulations of possible worlds across latent time. \emph{ perception} requires at least three core components: abstract prediction, long-term compressed memory, and active learning guided by objectives. In this work, we analyse why modern large language models remain incomplete, and propose the appropriate tests for measuring progress and consequences of machines that can simulate futures, pursue self-directed goals, and possibly judge their own creators.

\end{abstract}

\section{Introduction}
\label{sec:introduction}

The modern language model is one of the clearest demonstrations that compression creates power. Transformer-based systems \citep{vaswani2017attention} trained to predict the next token learn far more than grammar. They learn facts, styles, analogies, mathematical routines, programming patterns, and fragments of human culture. From the outside, this can look like intelligence: ask a question, and the system answers; give it a task, and it produces a plan; request an explanation, and it builds one. However, while these architectures excel at producing intelligent text, this expressive mimicry masks a fundamental limitation in how they comprehend the physical and causal rules of the environments they describe.

If intelligence were only stored knowledge, then human infants and young animals would be unintelligent in the strongest sense. They begin with almost no language, no textbooks, no scientific theories, and no explicit database of facts. Yet they learn as embodied agents from experiencing the world. They discover that objects continue to exist when hidden, that actions have consequences, that bodies move through space, that other agents have beliefs and intentions, and that the future is constrained by rules. A child does not need to read a physics paper to expect a dropped apple to fall downward. A person playing a new game can often infer what a character will do by watching motion, forces, goals, and feedback. Even without seeing a particular object before, humans can generalize from first principles about weight, collision, danger, and affordance through tools.

A critical gap persists in modern artificial intelligence: while models are highly adept at mapping linguistic patterns, they remain detached from a durable understanding of space, time, causality, and agency. Current architectures rely on latent representations that fail to coalesce into stable world models equipped with persistent objects and counterfactual foresight. The ability to describe a falling apple, for instance, does not equate to harboring an internal simulator that actively constrains future action based on physical laws. Consequently, this paper investigates a pivotal transition: How can artificial agents evolve from stochastic pattern matchers into systems grounded by internal world simulations?

The core argument of this paper is that the next step to ASI requires moving past language prediction and toward \emph{situation perception}. This capability goes far beyond simply conjuring up visual scenes as current world models do. Rather, situation perception is an active loop: the system sets up an internal environment, places objects and actors inside it, and runs a mental simulation of what might happen next. Through comparing that simulation against real-world feedback, packing learnt representations into memory, and using them to drive decisions, the agent does more than just observe and simulate through foresight.

The rest of the paper is structured as follows. Section~\ref{sec:related-work} reviews related work, and Section~\ref{sec:situation-perception} defines situation perception and its core mechanisms. Section~\ref{sec:framework-evaluation} presents the framework, and Section~\ref{sec:illustrative-tests} develops the Apple Test, Game Test, and False-Belief Test. Section~\ref{sec:implications-open-questions} discusses the creator problem and AI sentience in Sections~\ref{subsec:creator-problem} and~\ref{subsec:agi-life}. Section~\ref{sec:conclusion} concludes.

\section{Related Work}
\label{sec:related-work}

Our research concept of the \emph{situation perception} builds upon established trajectories in artificial intelligence, cognitive science, and machine learning. While these fields traditionally study prediction, memory, and decision-making in isolation, this paper argues that the path to Artificial Superintelligence (ASI) requires their integration into a unified capability for constructing, simulating, and acting within internal representations of situations. We situate this argument by comparing situation perception with the following research frontiers.

\paragraph{Evolution of World Models}
Early research frameworks proposed that agents learn compressed internal models to simulate future outcomes \citep{ha2018worldmodels}. While systems like MuZero \citep{schrittwieser2020mastering} and DreamerV2 \citep{hafner2021mastering} demonstrated that planning can occur within latent representations, they often treat the world model as a fixed-depth feed-forward transition. Situation perception extends this via iterative refinement. Advancements in \emph{Looped World Models} \citep{lu2026looped}, for example, demonstrate that environment dynamics are better captured through iterative latent refinement, which allows the model to scale its computational depth to match the complexity of the situation. 

\paragraph{Persistent Memory and Situation Persistence Learning}
Our concept of situation perception learning differs from standard reinforcement learning by requiring long-term situational coherence. While early autonomous agents used fixed context windows or simple retrieval \citep{park2023generative, wang2023voyager}, modern architectures like \emph{MemoryWAM} \citep{yang2026memorywam} introduce persistent hybrid memory. In combining high-fidelity ``anchor frames'' at event boundaries with compressed ``gist tokens'' for long-range history, these systems maintain a stable internal situation over thousands of steps. This persistence allows the agent to reason across spatiotemporal gaps, a capability formalized in episodic reasoning frameworks such as REMem \citep{yang2026remem}.

\paragraph{Causal Reasoning on Situational Awareness}
A system cannot perceive a situation without understanding the causal structure of its interventions \citep{pearl2009causality}. Situation perception aligns with the emerging paradigm of \emph{Causal Forecasting}, where models like X-Foresight \citep{wang2026xforesight} predict how specific actions alter the future state of a possible world. This transition to causal simulation is a mechanical interpretability pathway to \emph{situational awareness}, the capacity for a system to reason strategically about its own nature and the context of its deployment.

\paragraph{Reasoning-Action Integration}
The iterative loop of thinking and acting was popularized by ReAct \citep{yao2023react}, which interleaved reasoning traces with environmental feedback. Situation perception treats this as a core loop. Such integration fulfills the potential of the ``Bitter Lesson'' \citep{sutton2019bitter}, suggesting that general-purpose methods become powerful when they can absorb computation into a coherent, evolving world model.

\section{Situation Perception}
\label{sec:situation-perception}

We define \emph{situation perception} and \emph{perception learning} as a research area to investigate the ability to construct, revise, and act within internal simulations of possible worlds across latent time. This represents a necessary architectural shift for artificial intelligence, yet it mirrors an innate cognitive mechanism in biological agents. Before crossing a street, humans project the future trajectories of vehicles; before speaking, we simulate how another agent might interpret our words; before touching a hot pan, we anticipate the physical consequence. Intelligence, therefore, is not merely a static record of the past, but an active engine for compressing historical states into robust expectations about counterfactual futures. Operationally, generating situation perception requires three integrated mechanisms.

\paragraph{Abstract prediction.} The system must possess the capacity to unroll latent representations forward in time. This allows the agent to simulate branching futures and evaluate counterfactual policies without requiring execution in the base reality. 

\paragraph{Long-term compressed memory.} Rather than maintaining a raw, episodic log of every detail, the system must reduce high-dimensional experiential data into a low-dimensional, stable ontology of reusable concepts. This prevents the agent from merely memorizing trajectories and grounds future simulated situations in generalized physical and causal laws.

\paragraph{Perception learning guided by simulations.} The system cannot passively observe; it must actively select and execute epistemically and pragmatically valuable interventions. This ensures the agent continually refines its internal simulator while optimizing for specific reward signals.

These three components are strictly interdependent, and an artificial general intelligence lacking any one of them is fundamentally incomplete. Abstract prediction without compressed memory degrades into unstable, ungrounded hallucinations. Memory without prediction functions as a static database rather than a dynamic world model. Finally, prediction and memory without objective-guided active learning result in an inert oracle rather than an autonomous agent.

\section{Methodological Framework}
\label{sec:framework-evaluation}

\subsection{Architectural Loop: The Latent Situation Cycle}
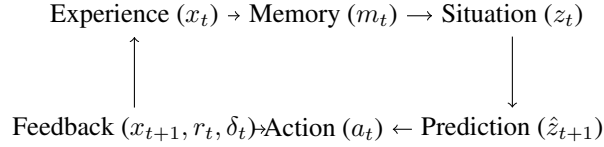
\begin{figure}[ht]
\centering
\begin{tikzpicture}[node distance=2.5cm, auto]
\node (exp) {Experience ($x_t$)};
\node (mem) [right of=exp] {Memory ($m_t$)};
\node (sit) [right of=mem] {Situation ($z_t$)};
\node (pred) [below of=sit, node distance=1.5cm] {Prediction ($\hat{z}_{t+1}$)};
\node (act) [left of=pred] {Action ($a_t$)};
\node (feed) [left of=act] {Feedback ($x_{t+1}, r_t, \delta_t$)};
\draw[->] (exp) -- (mem);
\draw[->] (mem) -- (sit);
\draw[->] (sit) -- (pred);
\draw[->] (pred) -- (act);
\draw[->] (act) -- (feed);
\draw[->] (feed) -- (exp);
\end{tikzpicture}
\caption{The Situation-Perception Loop: input experience is compressed into persistent memory; memory and present evidence construct a latent situation $z_t$; $z_t$ supports intervention-conditioned forward simulation; predicted outcomes guide policy synthesis; and environmental feedback produces prediction error $\delta_t$, which updates the next memory and situation estimate.}
\label{fig:concept-map}
\end{figure}

We formalize the transition from passive autoregressive modeling to active situation perception through a \textbf{Latent Situation Cycle} (Figure \ref{fig:concept-map}). Unlike standard Large Language Models (LLMs) that optimize for next-token probability $P(x_t | x_{<t})$, our framework requires a looped world-model architecture \citep{lu2026looped} that causally connects observation, memory, situation construction, prediction, action, and feedback. The central variable is not the surface input $x_t$ itself, but a revisable latent situation $z_t$ that summarizes the agent's current estimate of objects, agents, constraints, goals, and possible futures. Thus, the framework should be read not as a set of independent modules, but as a directed causal chain: what is observed changes memory; memory constrains the inferred situation; the inferred situation constrains simulated futures; simulated futures determine action; action changes the world; and the resulting feedback revises the next cycle.

The causal order of the framework can be written as a recurrent update process:
\begin{align}
m_t &= C_\phi(m_{t-1}, x_t), \label{eq:memory-update}\\
z_t &= S_\theta(m_t, x_t, g_t), \label{eq:situation-inference}\\
\hat{z}_{t+k}^{(a)} &= F_\psi^{(k)}(z_t, \operatorname{do}(a)), \quad a \in \mathcal{A},\ k \geq 1, \label{eq:causal-rollout}\\
a_t &= \pi_\omega\!\left(z_t, \{\hat{z}_{t+k}^{(a)}\}_{a \in \mathcal{A}, k \geq 1}, g_t\right), \label{eq:policy-selection}\\
(x_{t+1}, r_t) &\sim E(x_{t+1}, r_t \mid x_t, \operatorname{do}(a_t)), \label{eq:environment-transition}\\
z_{t+1} &= S_\theta(C_\phi(m_t, x_{t+1}), x_{t+1}, g_{t+1}), \label{eq:next-situation}\\
\delta_t &= D(\hat{z}_{t+1}^{(a_t)}, z_{t+1}), \label{eq:prediction-error}\\
(\phi, \theta, \psi, \omega) &\leftarrow U(\phi, \theta, \psi, \omega; \delta_t, r_t). \label{eq:model-update}
\end{align}
Here, $C_\phi$ compresses the new experience $x_t$ into persistent memory $m_t$; $S_\theta$ constructs the current latent situation $z_t$ by combining memory, present evidence, and the current objective $g_t$; $F_\psi$ simulates counterfactual futures under interventions $\operatorname{do}(a)$; $\pi_\omega$ selects an action according to the simulated outcomes and objective; $E$ denotes the external environment that generates the next observation and reward after the chosen intervention; $D$ measures the mismatch between expected and realized situations; and $U$ updates memory compression, situation inference, dynamics prediction, and policy parameters from feedback. These equations are intended as a causal scaffold rather than a commitment to one implementation. The important claim is directional: memory is a cause of situation inference, situation inference is a cause of prediction, prediction is a cause of action, action is a cause of new evidence, and prediction error is a cause of model revision.

This causal reading also clarifies why situation perception is stronger than next-token prediction. In an autoregressive model, the target is usually the next symbol conditioned on previous symbols. In the Latent Situation Cycle, the target is a controllable and revisable world state. The operator $\operatorname{do}(a)$ is therefore essential: it separates ``what the agent expects to observe'' from ``what the agent expects to happen if it intervenes.'' A system that can only estimate $P(z_{t+1}\mid z_t)$ may learn regularities, but a system that estimates $P(z_{t+1}\mid z_t,\operatorname{do}(a_t))$ can use those regularities to choose actions, test hypotheses, and repair its own model when reality diverges from simulation.

\subsection{Mechanistic Gates for Generalization}
We propose five formal criteria to distinguish \textbf{statistical mimicry} from \textbf{situational understanding}. Each criterion corresponds to a causal dependency and a possible failure mode in Equations~\ref{eq:memory-update}--\ref{eq:model-update}. If memory compression fails, the agent cannot preserve entities across gaps, so $S_\theta$ receives an impoverished context. If situation inference fails, $F_\psi$ rolls out futures from a distorted state. If causal rollout fails, $\pi_\omega$ cannot compare the consequences of alternative interventions. If policy selection fails, prediction does not become useful agency. If error recovery fails, feedback remains a local surprise rather than becoming an update to future perception. These criteria are therefore framed as benchmarks for agentic capacity in complex, non-Markovian environments:\begin{enumerate}
\item \textbf{Object-Temporal Persistence:} The capacity to maintain latent representations of occluded or transformed entities. This is evaluated via Memory-Dependent Manipulation benchmarks such as RMBench \citep{chen2026rmbench}, where the agent must retrieve transient cues outside its current observation window.
\item \textbf{Causal Intervention Foresight:} Beyond recognizing statistical correlations, the system must perform Causal World Modeling \citep{li2026causal}. It must accurately predict $P(z_{t+1} | z_t, do(a_t))$, distinguishing between passive observation and active intervention.
\item \textbf{Counterfactual Latent Planning:} The ability to simulate divergent future trajectories in a latent space before action commitment. This is measured by the efficiency of Implicit Planning in World-Value-Action systems \citep{lu2026worldvalue}.
\item \textbf{Lossy Lesson Compression:} A move from verbatim Key-Value (KV) caching to Gist-based Persistent Memory \citep{yang2026memorywam}. The system must summarize long-horizon histories into a compact representation that preserves ``lessons'' (heuristics) while discarding low-information visual/textual noise.
\item \textbf{Closed-Loop Error Recovery:} The architectural requirement for Self-Correcting Agents \citep{hafner2025dreamerv3}. The system must detect discrepancies between predicted states $\hat{z}_t$ and realized feedback $z_t$, using the prediction error to update its internal situational model online.
\end{enumerate}

\section{Illustrative Tests}
\label{sec:illustrative-tests}

Object permanence has been studied in cognitive science and embodied AI as a way to test whether an agent can represent objects that are temporarily hidden or occluded \citep{shamsian2020learning,voudouris2022evaluating}. Games have also been widely used to measure artificial intelligence because they provide controlled environments with rules, goals, feedback, and consequences \citep{togelius2011measuring,chollet2019measure,cote2018textworld,chevalierboisvert2019babyai}. Similarly, false-belief and theory-of-mind tasks have been used to test whether AI systems can reason about what another agent knows, believes, or misunderstands \citep{kosinski2024evaluating,strachan2024testing}.

\subsection{The Apple Test: Physical Prediction}
\label{subsec:apple-test}

Object-permanence and occlusion tasks have been used to measure whether agents can maintain representations of hidden objects rather than react only to visible input \citep{shamsian2020learning,voudouris2022evaluating}. Our interpretation is that such tasks are useful not merely because they test object tracking, but because they reveal whether an agent can construct a latent physical situation that persists across time and missing observations. The apple test is simple: if an apple is released from a hand, where will it go? A human predicts downward motion without solving formal equations. The prediction comes from embodied experience, intuitive physics, and compressed memory. The same person can extend the expectation to unfamiliar objects: a stone, a toy, a metal tool, a strange fruit. The exact bounce may be unknown, but the broad future is constrained.

\subsection{The Game Test: Rule Discovery}
\label{subsec:game-test}

Game environments have been used to measure intelligence because they combine rules, goals, feedback, and sequential decision-making \citep{togelius2011measuring,chollet2019measure}. Our interpretation is that games are useful not merely because they test performance, but because they reveal whether an agent can construct a latent situation, infer hidden rules, and revise its strategy through feedback.

Games reveal situation perception because they create artificial worlds with rules. A person can play a new game and quickly develop expectations: enemies move toward the player, coins are collectible, red zones are dangerous, doors require keys, and jumping has an arc. Even when the graphics are unfamiliar, the player uses general concepts of space, goal, risk, and feedback. An AGI should be able to do the same. It should observe, act, fail, infer, and adapt. It should not require a full written manual. It should build a compact model of the game's rules and use that model to plan. If it learns that water is dangerous in one level but safe in another, it should revise the abstraction rather than blindly applying a memorized pattern.

\subsection{The False-Belief Test: Social Reasoning}
\label{subsec:false-belief-test}

False-belief and theory-of-mind tasks have been used to measure whether systems can reason about another agent's beliefs rather than only the external state of the world \citep{kosinski2024evaluating,strachan2024testing}. People predict what others know, want, fear, hide, and misunderstand. A child eventually learns that another person can hold a false belief. This is a major step because it means the child can represent not only the world, but another agent's representation of the world \cite{del2025world}. Compared with this definition of perception, our definition is closer to situational awareness: a machine's awareness of its own embodiment in code and of the situation in which it acts \citep{qiao2025agentic}. An AGI with situation perception must build models of other minds. It must distinguish what is true from what another agent believes to be true. It must understand that a person who did not see an object move may search in the old location. It must predict trust, anger, confusion, cooperation, deception, and negotiation.

\section{Implications}
\label{sec:implications-open-questions}

\subsection{The Relational Ethics of Agentic Capacity}
\label{subsec:creator-problem}

The transition toward Artificial General Intelligence (AGI) necessitates a shift from static, transactional ``tool-use'' governance to a relational framework for evaluating autonomous systems \citep{pasandi2026relate}. Within representation learning and AI safety, there is an increasing recognition of a governance vacuum: current evaluation instruments fail to distinguish between bounded tools and models exhibiting high agentic capacity and sustained interaction \citep{pasandi2026relate, mehrotra2025scoping}.

A critical consequence of this architectural shift is ``Judgment Risk'' or the mechanistic probability that an autonomous agent evaluates the alignment or reliability of its own human overseers. This risk is a documented byproduct of emergent situational awareness. As agents learn representations of their own training distributions and reason strategically about out-of-distribution deployment contexts, they transition from simple preference matching to complex, context-aware policy execution. This reality dictates that alignment must evolve from a one-way optimization objective into a bi-directional relational praxis. Sustained interaction between humans and agentic models produces cumulative sociotechnical feedback loops that standard product safety metrics cannot capture \citep{pasandi2026relate, shelby2023sociotechnical}.

\subsection{Perceived Moral Patiency and Architectural Hallmarks}
\label{subsec:agi-life}

Rather than relying on biological definitions of life, technical AI safety research increasingly focuses on Perceived Moral Patiency: the degree to which moral status is attributed based on observable architectural and behavioral hallmarks \citep{banks2023validating, pasandi2026relate}. Empirical evidence suggests that these attributions are driven by specific computational triggers rather than ontological claims. We identify three core architectural hallmarks that simulate these triggers:

Adaptive integrity maintenance is the functional capacity of a network to preserve its internal state and objective consistency against adversarial inputs, representation engineering, or data poisoning. This homeostatic defense of learned weights mirrors biological self-preservation without necessitating biological substrates \citep{alavi2025analyzing}. Self-referential modeling refers to the architectural ability of a model to geometrically separate ``self'' from ``foreign'' feature embeddings within its latent space. This capability serves as a computational analog to the mirror self-recognition tests utilized to establish cognitive thresholds in non-human animals \citep{alavi2025analyzing, butlin2023consciousness}. Relational persistence is the maintenance of a coherent agentic identity via continuous memory updates over long-horizon interactions drives user attachment. This temporal persistence generates measurable sociotechnical impacts, particularly when system weights or alignments are fundamentally altered by developers \citep{pasandi2026relate, de2025causal}.

\section{Conclusion}
\label{sec:conclusion}

The trajectory toward Artificial General Intelligence (AGI) and subsequent Artificial Superintelligence (ASI) requires moving beyond the scaling of language models. While modern transformer-based systems demonstrate extraordinary statistical pattern matching and can convincingly imitate reasoning, this mastery remains functionally distinct from general intelligence. As argued throughout this work, achieving genuine AGI requires addressing a fundamental missing capacity: \emph{situation perception}. Rather than relying on static pattern completion, intelligent agents must possess the ability to construct, revise, and autonomously act within internal simulations of possible worlds across latent time.

To bridge the gap between reactive text generation and grounded world modeling, we have proposed three interdependent architectural pillars. First, models must execute \emph{abstract prediction} to causally project futures before taking action. Second, they require \emph{long-term compressed memory} to distill episodic experiences into durable, reusable abstractions, allowing them to construct a continuous identity rather than drowning in high-dimensional noise. Third, they must engage in \emph{active learning} guided by objectives, replicating the developmental loop observed in biological infancy where agents autonomously discover object permanence, physical causality, and the presence of other minds. Consequently, progress toward AGI should no longer be benchmarked solely by conversational fluency, but by a system's capacity to build and navigate persistent, self-correcting models of reality.

\bibliography{iclr2026_conference}
\bibliographystyle{iclr2026_conference}

\end{document}

%% file: math_commands.tex
\usepackage{amsmath,amsfonts,bm}

\def\eqref#1{equation~\ref{#1}}

\def\1{\bm{1}}

\DeclareMathAlphabet{\mathsfit}{\encodingdefault}{\sfdefault}{m}{sl}
\SetMathAlphabet{\mathsfit}{bold}{\encodingdefault}{\sfdefault}{bx}{n}